\documentclass[conference]{IEEEtran}
\usepackage[square, comma, sort&compress, numbers]{natbib}

\ifCLASSINFOpdf
\else
\fi

\usepackage{flushend}
\usepackage{graphicx}
\usepackage{float} 
\usepackage{subfigure}
\usepackage[]{caption2} 
\renewcommand{\figurename}{Fig.} 

\usepackage[table,xcdraw]{xcolor}
\usepackage{booktabs,graphicx}

\usepackage{authblk}
\usepackage{multirow}
\begin{document}
%
\title{Hand Gesture Recognition based on Radar Micro-Doppler Signature Envelopes}



%


\author{Moeness G. Amin\\
Center for Advanced Communications\\
Villanova University\\
Villanova, PA 19085, USA \\
moeness.amin@villanova.edu
\and
Zhengxin Zeng, Tao Shan\\
School of Information and Electronics\\
Beijing Institute of Technology\\
Beijing, China \\
\{3120140293,shantao\}@bit.edu.cn \and
}
\renewcommand*{\Affilfont}{\small\it} 
\renewcommand\Authands{ and }
\date{}



\maketitle

\makeatletter
\def\footnoterule{\kern-3\p@
  \hrule \@width 3.5in \kern 2.6\p@} 
\makeatother

\let\thefootnote\relax\footnotetext{The work of Mr. Zhengxin is funded by the International Graduate Exchange Program of Beijing Institute of Technology, and was performed while he was a Visiting Scholar at the Center for Advanced Communications, Villanova University}

\begin{abstract}
We introduce a simple but effective technique in automatic hand gesture recognition using radar. The proposed technique classifies hand gestures based on the envelopes of their micro-Doppler (MD) signatures. These envelopes capture the distinctions among different hand movements and their corresponding positive and negative Doppler frequencies that are generated during each gesture act. We detect the positive and negative frequency envelopes of MD separately, and form a feature vector of their augmentation. We use the $k$-nearest neighbor ($k$NN) classifier and Manhattan distance (L1) measure, in lieu of Euclidean distance (L2), so as not to diminish small but critical envelope values. It is shown that this method outperforms both low-dimension representation techniques based on principal component analysis (PCA) and sparse reconstruction using Gaussian-windowed Fourier dictionary, and can achieve very high classification rates. 
\end{abstract}


\IEEEpeerreviewmaketitle

\section{Introduction}
Radar systems assume an important role in several areas of our daily life, such as air traffic control, speed enforcement systems, and advanced driver assistance systems \cite{hopkin2017human,du2018vehicular,zolock2016use}. Recently, radar has also become of increased interest for indoor applications. In particular, human activity monitoring radar systems are rapidly evolving with applications that include gait recognition, fall motion detection for elderly care and aging-in-place technologies \cite{amin2017radar,amin2016radar}.\par
Over the past decade, much work has been done in human motion classifications which include daily activities of walking, kneeling, sitting, standing, bending, falling, etc. \cite{jokanovic2016radar,van2008feature,kim2015human,kim2016human,seyfiouglu2018deep,jokanovic2018fall,erol2018automatic,mobasseri2009time,seifert2018radar,gurbuz2016micro,seifertsubspace,le2018human,chen2018personnel}. Distinguishing among the different motions is viewed as an inter-class classification \cite{jokanovic2016radar,van2008feature,kim2015human,kim2016human,seyfiouglu2018deep,jokanovic2018fall,erol2018automatic}, whereas the intra-class classification amounts to identifying the different members of the same class, e.g., classifying normal and abnormal gaits \cite{mobasseri2009time,seifertsubspace,seifert2018radar,le2018human,gurbuz2016micro,chen2018personnel}. There are two main approaches of human motion classifications, namely those relying on handcrafted features that relate to human motion kinematics \cite{van2008feature,kim2015human,mobasseri2009time,seifert2018radar,gurbuz2016micro}, and others which are data driven and include low-dimension representations \cite{jokanovic2016radar,seifertsubspace}, frequency-warped cepstral analysis \cite{erol2018automatic}, and neural networks \cite{le2018human,kim2016human,seyfiouglu2018deep,jokanovic2018fall,chen2018personnel}.\par
In addition to classifying human motions, radars have been recently used for gesture recognition which is an important problem in a variety of applications that involve smart homes and human-machine interface for intelligent devices \cite{li2018sparsity,8367567,zhang2016dynamic,kim2016hand,zhang2018latern,li2018effect,yang2018sparsity,sakamoto2018hand,wang2016interacting}. The latter is considered vital in aiding the physically impaired who might be wheelchair confined or bed-ridden patients. The goal is to enable these individuals to be self-supported and independently functioning. In essence, automatic hand gesture recognition is poised to make our homes more user friendly and most efficient through the use of contactless radio frequency (RF) sensors that can identify different hand gestures for instrument and household appliance control. The most recent project Soli by Google for touchless interactions is a testament of this emerging technology \cite{wang2016interacting}. \par
The same approaches employed for classifying human daily activities can be applied for recognition of hand gestures using the electromagnetic (EM) sensing modality. However, there is an apparent difference between MD signatures of hand gestures and those associated with motion activities that involve human body. Depending on the experiment setup and radar data collection specs,  MD representations of hand gestures can be simple, limited to short time duration and small frequency bandwidth, and are mainly characterized by their confined power concentrations in the time-frequency domain. On the other hand, the MD signatures of body motions are intricate, of multi-component signals, span relatively longer time periods and assume higher Doppler frequencies.\par
In this paper, we present a method to discriminate five classes of dynamic hand gestures using radar MD sensor. These classes are swiping hand, hand rotation, flipping fingers, calling and snapping fingers. We begin with several types of hand motions, and use the canonical angle metric to assess the subspace similarities constructed from their respective time-frequency distributions \cite{jokanovic2017suitability}. Based on the results, we group these motions into five most dissimilar classes. Two MD features are extracted from the data spectrograms. They correspond to the positive and negative frequency envelopes of the hand gesture MD signatures. Only two envelopes implicitly capture, for each motion, the positive-negative frequency differences, the time alignments and misalignments of the peak positive and negative Doppler frequencies, and the signature extent and occupancy over the joint time and frequency variables. \par
We compare the proposed approach with that based on subspace decomposition methods, namely PCA \cite{jokanovic2016radar,seifertsubspace} and with the one using compressive sensing methods, namely sparse reconstruction employing Gaussian-windowed Fourier dictionary \cite{li2018sparsity}. In the latter, the classifier was applied to hand gesture data showing  signatures comprising rather detached power concentrated time-frequency regions.  The data collected from our radar sensing experiments demonstrate that our proposed method outperforms the above two methods, and achieve a classification accuracy higher than 96\%.\par
The remainder of this paper is organized as follows.  In Section II, we present the extraction method of MD signature envelopes and discusses the employed classifier. Section III describes the radar data collection and pre-processing of hand gestures. Section IV gives the experimental results based on the real data measurements. Section V is the conclusion of the paper.

\section{Hand Gesture Recognition Algorithm}
\subsection{Time-frequency Representations}
Hand gestures generate non-stationary radar back-scattering signals. Time-frequency representations (TFRs) are typically employed to analyze these signals in the joint-variable domains, revealing what is referred to as MD signatures.  A typical technique of TFRs is the spectrogram. For a discrete-time signal \(s(n)\) of length \(N\),  the spectrogram can be obtained by taking the short-time Fourier transform (STFT)
\begin{equation}\label{stft}
              S\left( {n,k} \right) = {\left| {\sum\limits_{m = 0}^{L - 1} {s(n + m)h(m){e^{ - j2\pi \frac{{mk}}{N}}}} } \right|^2}\
\end{equation}
where \(n=0,\cdots,N-1\) is the time index, $k=0,\cdots\,K-1$ is the discrete frequency index, and $L$ is the length of the window function $h(\cdot)$. The zero-frequency component is then shifted to the center of the spectrogram. It is noted that if the MD signal can be modeled as a sum of frequency modulated signals, then the signal parameters can be estimated using maximum likelihood techniques \cite{setlur2006analysis}. However, the MD signal of the hand gesture does not conform to this model and, as such, spectrograms will be used for feature extractions. It is also noted that the signal $s(n)$ in equation (\ref{stft}) is considered as a non-stationary deterministic signal rather than a random process \cite{amin1996minimum}.
\subsection{Extraction of the MD Signature Envelopes}
We select features specific to the nominal hand gesture local frequency behavior and power concentrations. These features are the positive and negative frequency envelopes in the spectrograms. The envelopes attempt to capture, among other things, the maximum positive frequency and negative frequencies, length of the event and its bandwidth, the relative emphases of the motion towards and away from the radar, i.e., positive and negative Doppler frequencies. In essence, the envelopes of the signal power concentration in the time-frequency domain may uniquely characterize the different hand motions.
The envelopes of the MD signature can be determined by an energy-based thresholding algorithm \cite{erol2017range}. First, the effective bandwidth of each gesture frequency spectrum is computed. This defines the maximum positive and negative Doppler frequencies. Second, the spectrogram $S(n,k)$ is divided into two parts, the positive frequency part and the negative frequency part. The corresponding energies of the two parts,   ${E_U}(n)$ and  ${E_L}(n)$,  at slow-time are computed separately as,

 \begin{equation}\label{uenergy}
              {E_U}\left( n \right) = \sum\limits_{k = 0}^{\frac{K}{2} - 1} {S{{\left( {n,k} \right)}^2}} ,  
               {E_L}\left( n \right) = \sum\limits_{k = \frac{K}{2}}^{K-1} {S{{\left( {n,k} \right)}^2}} 
\end{equation}
These energies are then scaled to define the respective thresholds, ${T_U}$ and ${T_L}$,  
\begin{equation}\label{uthreshold}
              {T_U}(n) = {E_U}(n)\cdot {\sigma_U}, {T_L}(n) = {E_L}(n)\cdot {\sigma_L}
\end{equation}
where ${\sigma_U}$ and ${\sigma_L}$ represent the scale factors, both are less than 1. These scalars can be chosen empirically, but an effective way for their selections is to maintain the ratio of the energy to the threshold values constant over all time samples. For the positive frequency envelope, this ratio can be computed by finding both values at the maximum positive Doppler frequency. Once the threshold is computed per equation (\ref{uthreshold}), the positive frequency envelope is then found by locating the Doppler frequency for which the spectrogram assumes equal or higher value.   Similar procedure can be followed for the negative frequency envelope.  The positive frequency envelope ${e_U}(n)$ and negative frequency envelope $e_L(n)$, are concatenated to form a long feature vector $e=[{e_U},{e_L}]$.
\subsection{Classifier}
We apply proper classifiers based on the envelope features extracted from the spectrograms. The $k$NN and Support vector Machine (SVM) are among the most commonly used classifiers in pattern recognition which are used in this paper. In particular, the $k$NN is a simple machine learning classification algorithm, where for each test sample, the algorithm calculates the distance to all the training samples, and selects the $k$ closest training samples. Classification is performed by assigning the label that is most frequent among these samples \cite{cover1967nearest}. Clearly, the best choice of $k$ would depend on the data.  In this work, $k$ is set to 1. Four different  distance metrics are considered, namely, the Euclidean distance,  the Manhattan distance \cite{wang2007improving}, the Earth Mover's distance (EMD) \cite{rubner2001earth} and the modified Hausdorff distance (MHD) \cite{dubuisson1994modified}.\par
SVM is a supervised learning algorithm  \cite{cortes1995support}. It exhibits clear advantages in nonlinear and high dimension problems.
\section{Hand Gesture Subclasses}
The data analyzed in this paper was collected in the Radar Imaging Lab at the Center for Advanced Communications, Villanova University. The radar system used in the experiment generates continuous wave, with carrier frequency and sampling rate equal to 25 GHz and 12.8 kHz, respectively. The radar was placed at the edge of a table. The gestures were performed approximately 20 cm away from radar at zero angle while the individual was sitting down using the dominant hand, and the arm remained fixed as much as possible during each gesture motion.\par
\begin{figure}[!htb]
\vspace{-0.1cm}
\setlength{\belowcaptionskip}{-0.1cm}
\begin{minipage}[b]{0.19\textwidth}
\includegraphics[width=0.8\textwidth]{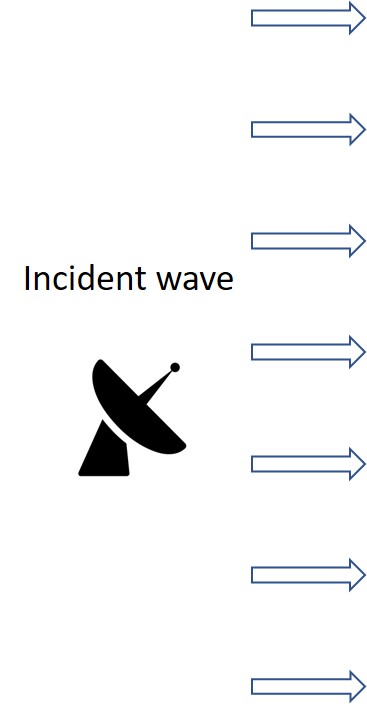} 
\end{minipage}
\begin{minipage}[b]{0.28\textwidth}
\subfigure[]{ 
\includegraphics[width=0.16\textwidth]{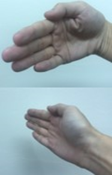}} 
 \subfigure[]{ 
  \includegraphics[width=0.16\textwidth]{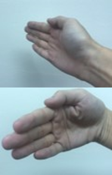}} 
   \subfigure[]{ 
  \includegraphics[width=0.16\textwidth]{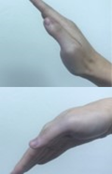}} 
   \subfigure[]{ 
  \includegraphics[width=0.16\textwidth]{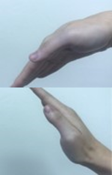}} 
   \subfigure[]{ 
  \includegraphics[width=0.16\textwidth]{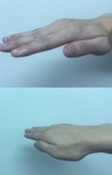}}\\ 
   \subfigure[]{ 
  \includegraphics[width=0.16\textwidth]{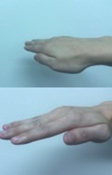}} 
   \subfigure[]{ 
  \includegraphics[width=0.16\textwidth]{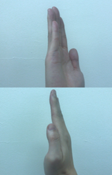}} 
   \subfigure[]{ 
  \includegraphics[width=0.16\textwidth]{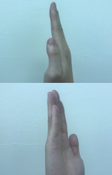}} 
   \subfigure[]{ 
  \includegraphics[width=0.16\textwidth]{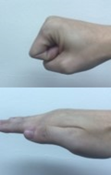}} 
   \subfigure[]{ 
  \includegraphics[width=0.16\textwidth]{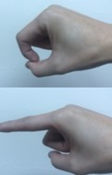}} \\
   \subfigure[]{ 
  \includegraphics[width=0.16\textwidth]{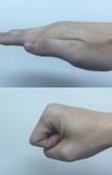}} 
   \subfigure[]{ 
  \includegraphics[width=0.16\textwidth]{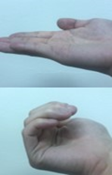}} 
   \subfigure[]{ 
  \includegraphics[width=0.16\textwidth]{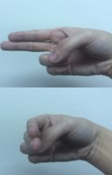}} 
   \subfigure[]{ 
  \includegraphics[width=0.16\textwidth]{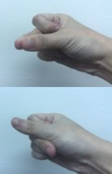}} 
   \subfigure[]{ 
  \includegraphics[width=0.16\textwidth]{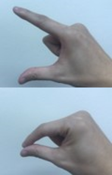}} 
\end{minipage}
  \caption{ Illustrations of 15 different hand gestures.} 
\label{photo}
\end{figure}

As depicted in Fig.\ref{photo}. The following 15 hand gestures were conducted: (a) Swiping hand from left to right, (b) Swiping hand from right to left, (c) Swiping hand from up to down, (d) Swiping hand from down to up, (e) Horizontal rotating hand clockwise, (f) Horizontal rotating hand counterclockwise, (g) Vertical rotating hand clockwise, (h) Vertical rotating hand counterclockwise, (i) Opening hand, (j) Flipping fingers, (k) Clenching hand, (l) Calling,  (m) Swipe left with two fingers, (n) Snapping fingers, (o) Pinching index. Four persons participated in the experiment. Each hand gesture was recorded over 8 seconds to generate one data segment. The recording was repeated for 5 times. Each data segment contained 2 or 3 individual hand motions, and a 1 second time window is applied to capture the individual motions. As such, repetitive motions and associated duty cycles were not considered in classifications. In total, 755 segments of data for 15 hand gestures were generated.\par

\begin{figure}[!htb]
\setlength{\belowcaptionskip}{-0.5cm}
\subfigure[]{ 
\includegraphics[width=0.11\textwidth]{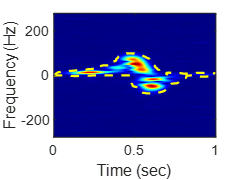}} 
 \subfigure[]{ 
  \includegraphics[width=0.11\textwidth]{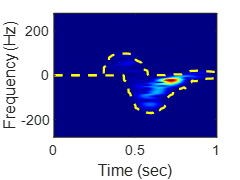}} 
   \subfigure[]{ 
  \includegraphics[width=0.11\textwidth]{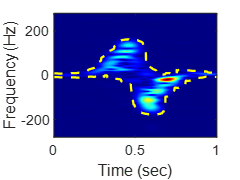}} 
   \subfigure[]{ 
  \includegraphics[width=0.11\textwidth]{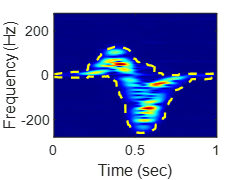}}
  \vspace{-2.5ex}\\
   \subfigure[]{ 
  \includegraphics[width=0.11\textwidth]{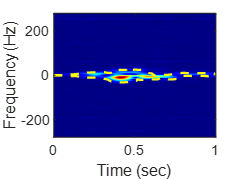}}
   \subfigure[]{ 
  \includegraphics[width=0.11\textwidth]{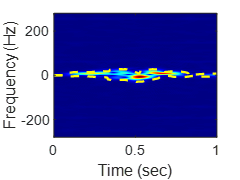}} 
   \subfigure[]{ 
  \includegraphics[width=0.11\textwidth]{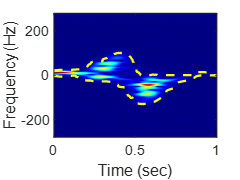}} 
   \subfigure[]{ 
  \includegraphics[width=0.11\textwidth]{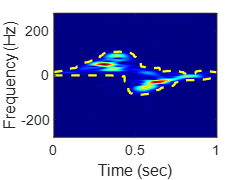}}
  \vspace{-2.5ex}\\
   \subfigure[]{ 
  \includegraphics[width=0.11\textwidth]{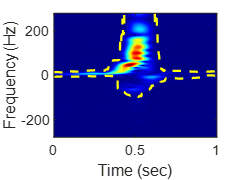}} 
   \subfigure[]{ 
  \includegraphics[width=0.11\textwidth]{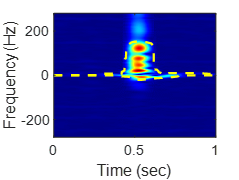}} 
  \vspace{-2.5ex}\\
   \subfigure[]{ 
  \includegraphics[width=0.11\textwidth]{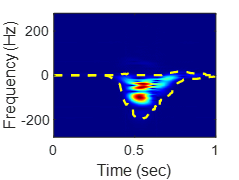}} 
   \subfigure[]{ 
  \includegraphics[width=0.11\textwidth]{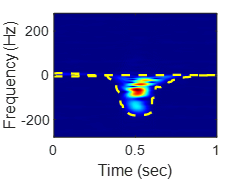}} 
   \subfigure[]{ 
  \includegraphics[width=0.11\textwidth]{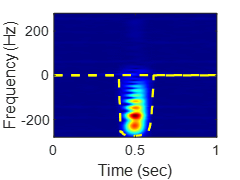}} 
  \vspace{-2.5ex}\\
   \subfigure[]{ 
  \includegraphics[width=0.11\textwidth]{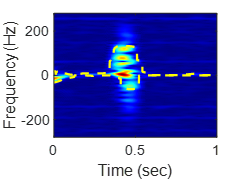}} 
   \subfigure[]{ 
  \includegraphics[width=0.11\textwidth]{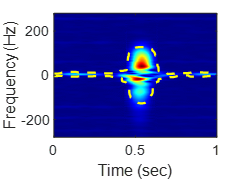}} 
  \caption{ Spectrograms and corresponding envelopes of 15 different hand gestures.} 
\label{spectrograms}
\end{figure}

Fig. \ref{spectrograms} shows examples of spectrograms and the corresponding envelopes for different hand gestures. The employed sliding window $h(\cdot)$ is rectangular with length $L=$2048 (0.16 $s$), and $K$ is set to 4096.
It is clear that the envelopes can well capture the salient features of the respective spectrograms.  It is also evident that the MD characteristics of the spectrograms are in agreement and consistent with each hand motion kinematics. For example, for the hand gesture `Swiping hand', the hand moves closer to the radar at the beginning which causes the positive frequency, and then moves away from the radar which induces the negative frequency. \par

Observing the spectrograms in Fig. \ref{spectrograms}, it is noticeable that similar motions generate similar signatures. To mathematically confirm these resemblances, we consider sub-grouping the 15 hand gestures using the Canonical correlation measure \cite{jokanovic2017suitability}. In this case, the spectrograms are converted to gray-scale images with the size $100\times 100$, and then vectorized with the size $1\times 10000$.   \par
Define matrix $X$ contains $M$ vectorized images ${S_i}, i=1,\cdots,M$ of a specific hand gesture,
\begin{equation}\label{classmatrix}
            X = [{x_1}|{x_2}| \cdots |{x_M}]
\end{equation}
\par
The $d$-dimensional subspace of a specific hand gesture can be obtained by taking PCA of $X$ \cite{jolliffe2011principal}. Suppose $\Phi_1$ and $\Phi_2$ are two $d$-dimensional linear subspaces, the canonical correlations of the two subspaces are the cosines of principal angles, and are defined as \cite{kim2007discriminative},
\begin{equation}\label{angle}
            \cos {\theta _i} = \mathop {\max }\limits_{{u_i} \in {\Phi _1}} \mathop {\max }\limits_{{v_i} \in {\Phi _2}} u_i^T{v_i}
\end{equation}
subject to  $||u||=||v||=1, {u_i}^Tu_j={v_i}^Tv_j=0, i\ne j$.  Let $U$ and $V$ denote unitary orthogonal bases for two subspaces,  $\Phi_1$ and $\Phi_2$. The singular value decomposition (SVD) of $U^TV$ is
\begin{equation}\label{svd}
            U^TV=P\Lambda Q
\end{equation}
\par
The canonical correlations are the singular values $\Lambda$, i.e., $\cos (\theta_i)=\lambda_i, i=1,\cdots,d$. The minimum angle is used to measure the closeness of two subspaces. Table \ref{canonical} shows the canonical correlations coefficients, from which we can clearly witnessed the similarities between the different hand gestures; larger coefficient indicates the two hand gestures are more similar. The red color of the table indicates the coefficient exceeds 0.9, and the yellow part means the coefficient is over 0.85. According to the Table numbers, we group the 15 hand gestures into 5 Class. Class I is the gesture `Swiping hand' which contains motions (a), (b), (c) and (d). Class II represents the gestures `Hand rotation' which contains motions (e), (f), (g) and (h). The gesture `Flipping fingers', which involves motions (i) and (j), makes Class III. Class IV is the gesture `Calling', which has motions (k), (l) and (m). The last Class V is the gesture `Snapping fingers'; it has motions (n) and (o). It is important to note that other similarity measures \cite{mitchell2010image} can be applied, in lieu of the canonical correlation. However, we found the canonical correlation most consistent with the visual similarities.

\begin{table}[!htb]
\setlength{\belowcaptionskip}{5pt}
\caption{ \sc Canonical Correlations Coefficients}
\resizebox{0.49\textwidth}{!}{
\begin{tabular}{|c|c|c|c|c|c|c|c|c|c|c|c|c|c|c|}
\hline
 & \textbf{b} & \textbf{c} & \textbf{d} & \textbf{e} & \textbf{f} & \textbf{g} & \textbf{h} & \textbf{i} & \textbf{j} & \textbf{k} & \textbf{l} & \textbf{m} & \textbf{n} & \textbf{o} \\ \hline
\textbf{a} & 0.79 & 0.83 & \cellcolor[HTML]{FF0000}{\color[HTML]{000000} 0.91} & 0.70 & 0.75 & 0.79 & 0.84 & 0.69 & 0.66 & 0.78 & 0.77 & 0.76 & 0.77 & 0.81 \\ \hline
\textbf{b} & 0 & \cellcolor[HTML]{FF0000}0.92 & 0.80 & 0.70 & 0.68 & 0.82 & 0.82 & 0.65 & 0.61 & 0.78 & 0.82 & 0.83 & 0.73 & 0.60 \\ \hline
\textbf{c} & 0 & 0 & 0.76 & 0.64 & 0.59 & \cellcolor[HTML]{F8FF00}0.85 & \cellcolor[HTML]{F8FF00}0.88 & 0.72 & 0.65 & 0.80 & 0.80 & 0.82 & 0.76 & 0.69 \\ \hline
\textbf{d} & 0 & 0 & 0 & 0.61 & 0.68 & 0.81 & 0.75 & 0.57 & 0.55 & 0.78 & 0.67 & 0.60 & 0.63 & 0.64 \\ \hline
\textbf{e} & 0 & 0 & 0 & 0 & \cellcolor[HTML]{F8FF00}0.86 & 0.70 & 0.75 & 0.59 & 0.66 & 0.56 & 0.72 & 0.66 & 0.72 & 0.71 \\ \hline
\textbf{f} & 0 & 0 & 0 & 0 & 0 & 0.78 & 0.83 & 0.70 & 0.70 & 0.67 & 0.73 & 0.70 & 0.78 & 0.79 \\ \hline
\textbf{g} & 0 & 0 & 0 & 0 & 0 & 0 & \cellcolor[HTML]{F8FF00}0.85 & 0.67 & 0.67 & 0.78 & 0.66 & 0.71 & 0.74 & 0.73 \\ \hline
\textbf{h} & 0 & 0 & 0 & 0 & 0 & 0 & 0 & 0.55 & 0.60 & 0.72 & 0.67 & 0.61 & 0.71 & 0.71 \\ \hline
\textbf{i} & 0 & 0 & 0 & 0 & 0 & 0 & 0 & 0 & \cellcolor[HTML]{F8FF00}0.87 & 0.75 & 0.61 & 0.67 & 0.76 & 0.74 \\ \hline
\textbf{j} & 0 & 0 & 0 & 0 & 0 & 0 & 0 & 0 & 0 & 0.68 & 0.61 & 0.68 & 0.83 & 0.73 \\ \hline
\textbf{k} & 0 & 0 & 0 & 0 & 0 & 0 & 0 & 0 & 0 & 0 & \cellcolor[HTML]{FF0000}0.94 & \cellcolor[HTML]{FF0000}0.94 & 0.83 & 0.76 \\ \hline
\textbf{l} & 0 & 0 & 0 & 0 & 0 & 0 & 0 & 0 & 0 & 0 & 0 & \cellcolor[HTML]{FF0000}0.93 & 0.73 & 0.66 \\ \hline
\textbf{m} & 0 & 0 & 0 & 0 & 0 & 0 & 0 & 0 & 0 & 0 & 0 & 0 & 0.77 & 0.63 \\ \hline
\textbf{n} & 0 & 0 & 0 & 0 & 0 & 0 & 0 & 0 & 0 & 0 & 0 & 0 & 0 & 0.82 \\ \hline
\end{tabular}}
\label{canonical}
\vspace{-0.5cm}
\end{table}
\section{Experimental Results}
In this section, all 755 data segments are used to validate the proposed method where 70\% of the data are used for training and 30\% for testing. The classification results are obtained by 1000 Monte Carlo trials. Three different automatic hand gesture approaches are compared with the proposed method. These are:  1) the empirical feature extraction method \cite{zhang2016dynamic}; 2) the PCA-based method \cite{seifertsubspace}; 3) the sparse reconstruction-based method \cite{li2018sparsity}.

\subsection{Empirical Feature Extraction Method}
\begin{figure}[!htp]
\vspace{-0.2cm}
\setlength{\belowcaptionskip}{-0.2cm}
\centering
\begin{minipage}[b]{0.24\textwidth}
\includegraphics[width=1\textwidth]{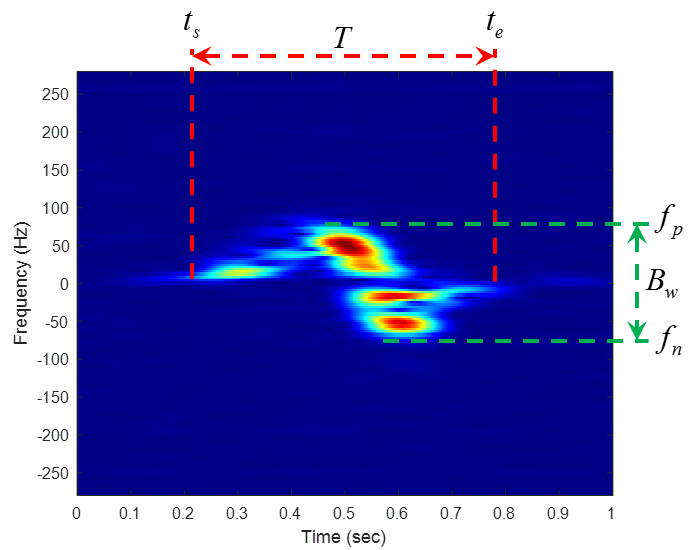}
\caption{Empirical feature extraction.}
\label{exspec}
\end{minipage}
\begin{minipage}[b]{0.24\textwidth}
\includegraphics[width=1\textwidth]{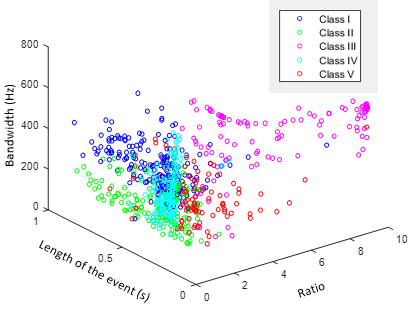}
\caption{Scatter plot of three extracted empirical features.}
\label{scatter}
\end{minipage}
\end{figure}

Three empirical features are extracted from the spectrograms to describe the hand gestures motions, namely the length of the event, the ratio of positive-negative frequency and the signal bandwidth. Fig. \ref{exspec} is an example showing these handcrafted features.\par
\textit{1) Length of the event $T$:} This describes the effective time duration to perform each hand gesture,
\begin{equation}\label{timelength}
            T=t_e-t_s
\end{equation}
where $t_s$ and $t_e$ represent the start time and the end time of a single hand gesture, respectively.
\par\textit{2) Ratio of positive-to-negative peak frequencies $R$:} This feature is obtained by finding ratio of the maximum positive frequency value, $f_p$, and maximum negative frequency value, $f_n$,
\begin{equation}\label{ratio}
            R=\left| {\frac{{{f_p}}}{{{f_n}}}} \right|
\end{equation}
where $|\cdot|$ is the absolute function.
\par\textit{3) Bandwidth $B_w$:} This is a measure of the the signal effective width,
\begin{equation}\label{bandwidth}
            B_w=|f_p|+|f_n|
\end{equation}

The scatter plot of the above extracted features is shown in Fig. \ref{scatter}. Table \ref{e} depicts the nominal behavior of these values over the different classes considered.  When using $k$NN-L1 as the classifier, the recognition accuracy based on these features is only 68\% with the confusion matrix shown in Table \ref{efcon}.


\begin{table}[!htb]
\centering
\vspace{-0.4cm}
\setlength{\belowcaptionskip}{5pt}
\caption{ \sc Nominal Behavior of Empirical Features over Different Classes}
\begin{tabular}{|c|c|c|c|}
\hline
\multicolumn{1}{|l|}{\multirow{2}{*}{}} & \multicolumn{3}{c|}{\textbf{Empirical features}}    \\ \cline{2-4} 
\multicolumn{1}{|l|}{}                  & \textbf{Time} & \textbf{Ratio} & \textbf{Bandwidth} \\ \hline
\textbf{Class I}                        & large         & moderate       & moderate           \\ \hline
\textbf{Class II}                       & large         & moderate       & small              \\ \hline
\textbf{Class III}                      & small         & large          & large              \\ \hline
\textbf{Class IV}                       & small         & small          & moderate           \\ \hline
\textbf{Class V}                        & small         & moderate       & moderate           \\ \hline
\end{tabular}
\vspace{-0.2cm}
\label{e}
\end{table}

\begin{table}[!htb]
\centering
\vspace{-0.4cm}
\setlength{\belowcaptionskip}{5pt}
\caption{ \sc Confusion Matrix Yielded by Empirical Feature Extraction Method}
\begin{tabular}{c|ccccc}
\hline
 & \textbf{I} & \textbf{II} & \textbf{III} & \textbf{IV} & \textbf{V} \\ \hline
\textbf{I} & 66.79\% & 13.80\%  & 4.08\%  & 9.18\%  & 6.15\%  \\
\textbf{II} & 20.04\%  & 64.65\%  & 3.53\%  & 4.88\%  & 6.90\%  \\
\textbf{III} & 9.94\%  & 5.59\%  & 76.53\%  & 0.03\%  & 7.91\%  \\
\textbf{IV} & 19.04\%  & 6.74\%  & 0.65\%  & 71.79\%  & 1.78\%  \\
\textbf{V} & 12.03\%  & 10.96\%  & 12.28\%  & 11.59\%  & 53.14\%  \\ \hline
\end{tabular}
\vspace{-0.46cm}
\label{efcon}
\end{table}

\subsection{Proposed Envelope-based Method}
As discussed in Section II, the extracted envelopes are fed into the $k$NN classifier, with different distance measures, and the SVM classifier. The recognition accuracy is presented in Table \ref{envacc}. It is clear that the $k$NN classifier based on L1 distance achieves the highest accuracy, over 96\%, followed by those employing the modified Hausdorff distance and the Euclidean distance. Different from other distances, the L1 distance attempts to properly account for small envelope values. The confusion matrix of the $k$NN classifier based on the L1 distance is shown in Table \ref{envcon}, from which we can observe that Class III and Class IV are most distinguishable, with an accuracy over 98\%.

\begin{table}[!htb]
\centering
\vspace{-0.2cm}
\setlength{\belowcaptionskip}{5pt}
\caption{ \sc Recognition Accuracy with Different Types of Classifier}
\begin{tabular}{c c}
\hline
 & \textbf{Accuracy }\\ \hline
SVM & 83.07\% \\ 
$k$NN-L1 & 95.23\% \\ 
$k$NN-L2 & 93.87\% \\ 
$k$NN-EMD & 81.51\% \\ 
$k$NN-MHD & 93.95\% \\ \hline
\end{tabular}
\vspace{-0.4cm}
\label{envacc}
\end{table}

\begin{table}[!htb]
\centering
\setlength{\belowcaptionskip}{5pt}
\caption{ \sc Confusion Matrix Yielded by Envelope Method based on $k$NN-L1 Classifier}
\begin{tabular}{c|ccccc}
\hline
 & \textbf{I} & \textbf{II} & \textbf{III} & \textbf{IV} & \textbf{V} \\ \hline
\textbf{I} & 95.23\% & 3.17\% & 0.14\%& 1.46\% & 0\\\
\textbf{II} & 3.03\%& 95.39\% & 0.01\% & 0.06\% & 1.51\% \\
\textbf{III} & 0.07\% & 0 & 99.01\%& 0.28\% & 0.64\% \\
\textbf{IV} & 0.61\%& 0 & 1.16\%& 98.21\% & 0.02\%\\
\textbf{V} & 0 & 2.31\% &2.61\% & 2.83\% & 92.25\% \\ \hline
\end{tabular}
\label{envcon}
\vspace{-0.2cm}
\end{table}

\subsection{PCA-based Method}
\begin{figure}[!htp]
\vspace{-0.4cm}
\centering
\includegraphics[width=0.25\textwidth]{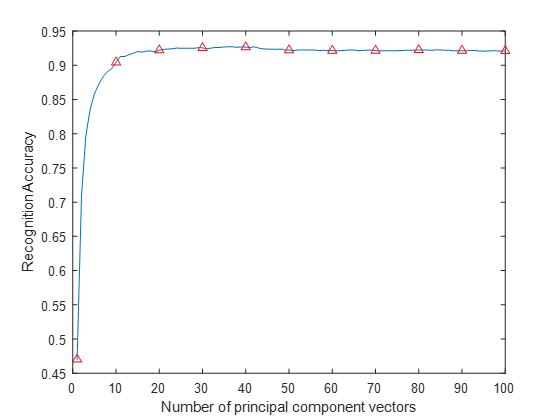}
\caption{Performance of PCA with different number of principal components.}
\label{PCA}
\vspace{-0.3cm}
\end{figure}

For the PCA-based method, each sample represents a spectrogram image of $100\times 100$ pixels. The number of principal components $d$ is determined by the eigenvalues. Fig. \ref{PCA} shows how the classification accuracy changes with $d$, with the recognition rate increases as $d$ increases. However, there is no significant improvement of the recognition accuracy past $d = 30$. Table \ref{pcacon} is the confusion matrix using $30$ eigenvalues. Although the PCA method can achieve an overall accuracy of 92.71\%, it is clearly outperformed by the proposed method.  

\begin{table}[]
\centering
\setlength{\belowcaptionskip}{5pt}
\caption{ \sc Confusion Matrix Yielded by PCA-based Method with $d=30$}
\begin{tabular}{c|ccccc}
\hline
 & \textbf{I} & \textbf{II} & \textbf{III} & \textbf{IV} & \textbf{V} \\ \hline
\textbf{I} & 89.50\% & 3.02\%& 0.67\%& 6.80\% & 0.01\%\\
\textbf{II} & 2.92\%& 94.83\%& 0 & 1.45\%& 0.80\%\\
\textbf{III} & 2.85\%& 1.23\%& 94.42\%& 0 & 1.50\%\\
\textbf{IV} & 5.24\% & 0.25\%& 1.37\% & 93.14\% & 0\\
\textbf{V} & 3.24\% & 8.14\% & 5.03\%& 1.83\%& 81.76\% \\ \hline
\end{tabular}
\label{pcacon}
\vspace{-0.4cm}
\end{table}

\subsection{Sparsity-based Method}
\begin{figure}[!htb]
\subfigure[]{ 
\includegraphics[width=0.11\textwidth]{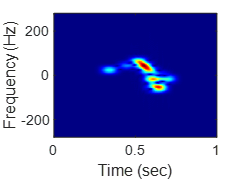}} 
 \subfigure[]{ 
  \includegraphics[width=0.11\textwidth]{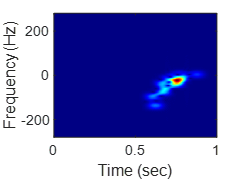}} 
   \subfigure[]{ 
  \includegraphics[width=0.11\textwidth]{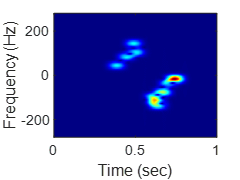}} 
   \subfigure[]{ 
  \includegraphics[width=0.11\textwidth]{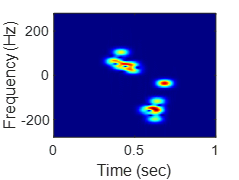}}
  \vspace{-2.5ex}\\
   \subfigure[]{ 
  \includegraphics[width=0.11\textwidth]{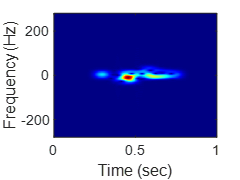}}
   \subfigure[]{ 
  \includegraphics[width=0.11\textwidth]{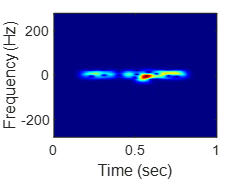}} 
   \subfigure[]{ 
  \includegraphics[width=0.11\textwidth]{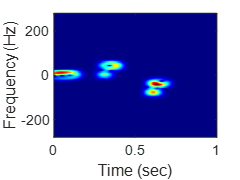}} 
   \subfigure[]{ 
  \includegraphics[width=0.11\textwidth]{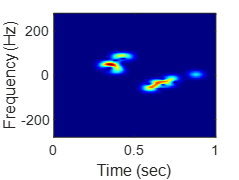}}
  \vspace{-2.5ex}\\
   \subfigure[]{ 
  \includegraphics[width=0.11\textwidth]{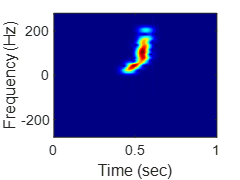}} 
   \subfigure[]{ 
  \includegraphics[width=0.11\textwidth]{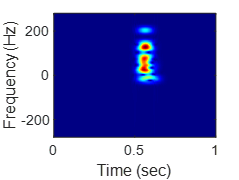}} 
  \vspace{-2.5ex}\\
   \subfigure[]{ 
  \includegraphics[width=0.11\textwidth]{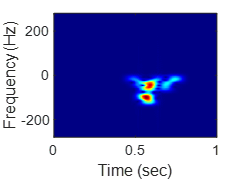}} 
   \subfigure[]{ 
  \includegraphics[width=0.11\textwidth]{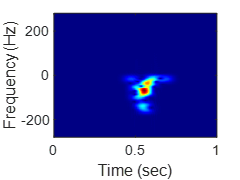}} 
   \subfigure[]{ 
  \includegraphics[width=0.11\textwidth]{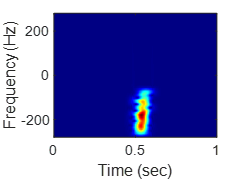}} 
  \vspace{-2.5ex}\\
   \subfigure[]{ 
  \includegraphics[width=0.11\textwidth]{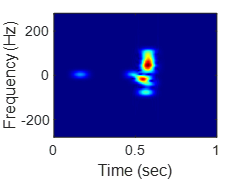}} 
   \subfigure[]{ 
  \includegraphics[width=0.11\textwidth]{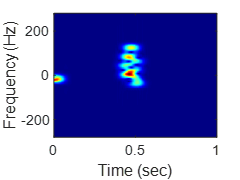}} 
  \caption{ Spectrograms of reconstructed signals with $P = 10$.} 
\label{sspectrograms}
\end{figure}

\begin{figure}[!htb]
\vspace{-0.4cm}
\subfigure[]{ 
\includegraphics[width=0.11\textwidth]{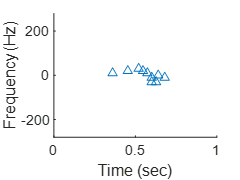}} 
 \subfigure[]{ 
  \includegraphics[width=0.11\textwidth]{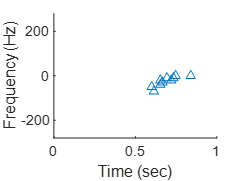}} 
   \subfigure[]{ 
  \includegraphics[width=0.11\textwidth]{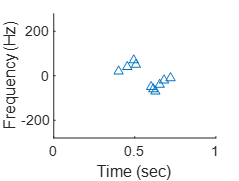}} 
   \subfigure[]{ 
  \includegraphics[width=0.11\textwidth]{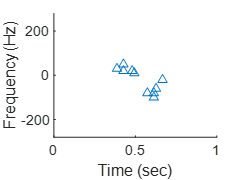}}
  \vspace{-2.5ex}\\
   \subfigure[]{ 
  \includegraphics[width=0.11\textwidth]{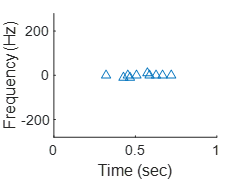}}
   \subfigure[]{ 
  \includegraphics[width=0.11\textwidth]{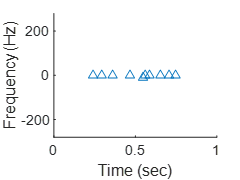}} 
   \subfigure[]{ 
  \includegraphics[width=0.11\textwidth]{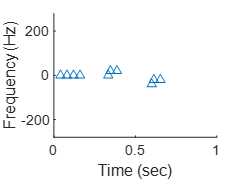}} 
   \subfigure[]{ 
  \includegraphics[width=0.11\textwidth]{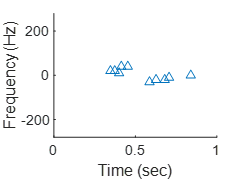}}
  \vspace{-2.5ex}\\
   \subfigure[]{ 
  \includegraphics[width=0.11\textwidth]{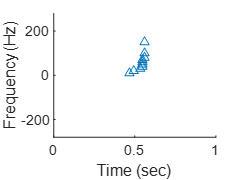}} 
   \subfigure[]{ 
  \includegraphics[width=0.11\textwidth]{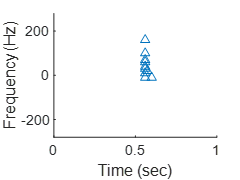}} 
  \vspace{-2.5ex}\\
   \subfigure[]{ 
  \includegraphics[width=0.11\textwidth]{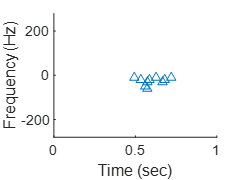}} 
   \subfigure[]{ 
  \includegraphics[width=0.11\textwidth]{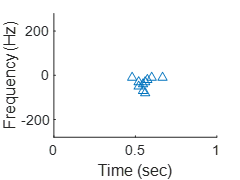}} 
   \subfigure[]{ 
  \includegraphics[width=0.11\textwidth]{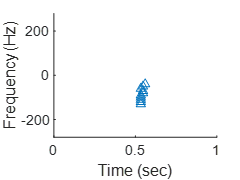}} 
  \vspace{-2.5ex}\\
   \subfigure[]{ 
  \includegraphics[width=0.11\textwidth]{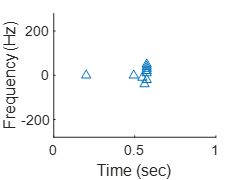}} 
   \subfigure[]{ 
  \includegraphics[width=0.11\textwidth]{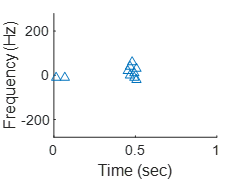}} 
  \caption{ Locations of time$-$frequency trajectories with $P=10$.} 
\label{sloc}
\vspace{-0.5cm}
\end{figure}

The features used for this method are the time-frequency trajectories. Details of the sparsity-based method can be found in \cite{li2018sparsity}. The trajectory consists of three parameters, namely the time-frequency position  $(t_i,f_i), i=1,\cdots,P$ and the intensity $A_i$, $P$ is the sparsity level that is set to 10 in this paper. Hence, each sample contains 30 features. The spectrograms of reconstructed signals and the $P$ locations of time-frequency trajectory are plotted in Fig. \ref{sspectrograms} and Fig. \ref{sloc}. In the training process, the $K$-means algorithm is used to cluster a central time-frequency trajectory \cite{kanungo2002efficient}. In the testing process, the $k$NN classifier based on the modified Hausdorff distance is applied to measure the distance between the testing samples and central time-frequency trajectories. The corresponding confusion matrix is presented in Table \ref{sparsecon}. The overall recognition accuracy was found to be only about 70\% when applied to our data.

\begin{table}[!htb]
\centering
\vspace{-0.2cm}
\setlength{\belowcaptionskip}{5pt}
\caption{ \sc Confusion Matrix Yielded by Sparsity-based Method}
\begin{tabular}{c|ccccc}
\hline
 & \textbf{I} & \textbf{II} & \textbf{III} & \textbf{IV} & \textbf{V} \\ \hline
\textbf{I} & 71.72\% & 11.36\%  & 1.45\%  & 11.74\%  & 3.73\%  \\
\textbf{II} & 10.95\%  & 81.29\%  & 2.57\%  & 0.28\%  & 4.91\%  \\
\textbf{III} & 7.40\%  & 2.10\%  & 83.63\%  & 0.69\%  & 6.18\%  \\
\textbf{IV} & 16.04\%  & 6.52\%  & 1.22\%  & 74.14\%  & 2.08\%  \\
\textbf{V} & 6.65\%  & 15.05\%  & 9.96\%  & 10.02\%  & 58.32\%  \\ \hline
\end{tabular}
\label{sparsecon}
\vspace{-0.2cm}
\end{table}
\section{Conclusion}
We introduced a simple but effective technique for automatic hand gesture recognition based on radar MD signature envelopes. No range information was incorporated. An energy-based thresholding algorithm was applied to separately extract the positive and negative frequency envelopes of the signal spectrogram. We used the canonical correlation coefficient to group 15 different hand gestures into five classes. The members of each class have close signature behaviors.  The extracted envelopes were concatenated and inputted to different types of classifiers. It was shown that the $k$NN classifier based on L1 distance achieves the highest accuracy and provided over 96 percent classification rate. The experimental results also demonstrated that the proposed method outperformed the lower dimensional PCA-based method, the sparsity-based approach using Gaussian-windowed Fourier dictionary, and existing techniques based on handcrafted features. The proposed approach does not use deep learning, though it might reach a classification performance close to that offered by Google Soli.



%
\footnotesize
\bibliographystyle{IEEEtran.bst}
\bibliography{IEEEabrv,refs.bib}

\end{document}